# Evolution of burst distribution in fiber bundle model

## I.V.Bezsudnov[1], A.A.Snarskii[2]

[1] Nauka - Service JSC, Moscow, Russia, Corresponding author, e-mail: biv@akuan.ru

[2] National Technical University of Ukraine "KPI", Dep. of General and Theoretical Physics, Kiev, Ukraine, e-mail: asnarskii@gmail.com

## Abstract

**We consider a fiber bundle model with a equal load sharing and uniformly distributed breakdown thresholds. A unified probability-theoretic approach was used to describe bundle under continuous and discrete load increase. It was shown that the ratio of distribution $D(\Delta)$ of avalanches of sizes $\Delta$ to the number of bundle load steps exactly corresponds to burst probability of size $\Delta$. Evolution of $\xi$ - power law distribution exponent of $D(\Delta)$ was studied as a function of loading step and bundle size. It was shown that $\xi$ does not depend on bundle size. In the numerical experiment, dependence $\xi$ on loading step was obtained with fiber bundle size up to $N = 10^{10}$. The regions of $\xi$ values constancy and the transition region were found. A distribution of fiber bursts on each loading step was recovered in the numerical simulations. It was shown that a change in the type of this distribution is the reason for evolution of $\xi$ values in the range of $-3 \div -5/2$.**
PACS numbers: 46.50.+a, 62.20.Mk, 81.05.Ni.

### Introduction

Different kinds of fiber bundle models (FBM) are widely and fruitfully used for simulation of behavior (breakdown) of composite materials. Numerous examples of such models can be found in the books [1-3], reviews [4-6] and references therein.

In modern times, the first work on FBM was written by Peirce in 1926 [7], it has not become irrelevant to this day. In 1945 Daniels published a paper [8] where the problem of fiber bundle burst was for the first time considered as mathematical statistics problem, rather than the subject of experimental science on materials. This concept and a variant of problem posed in [8] are the starting point in the investigations of FMB models up to now.

After the appearance of Sornette's paper in 1989 [9] this model attracted attention of many researchers, on the one hand, due to its simplicity, and on the other hand, due to variety of behavior demonstrated by systems described by FBM models. It is also remarkable that for a numerical simulation of a composite or other phenomenon described by FBM there is no need in extraordinary computational power.

Because of a random distribution of breaking thresholds in a fiber bundle, bursts due to loading occur irregularly. During each loading step $\Delta f$ a different number of fibers $\Delta$ is broken. The distribution $D(\Delta)$ - burst frequency- follows a power law distribution [10] $D(\Delta) \sim \Delta^\xi$, therefore, parameter $\xi$ characterizes bundle behavior (breakdown) in FBM model.

For the investigation of dependence $\xi(\Delta f)$ there were employed various probability-theoretic approaches, as well as numerical simulation, which yielded different values for $\xi$. Thus, at "low" $\Delta f$ ($\Delta f \to 0$) [10,11], the so-called "continuous" bundle loading, the theory predicted $\xi = -5/2$, whereas at "large" $\Delta f$ ($\Delta f > 1$) [12], the so-called "discrete" loading, $\xi = -3$.

The dependence $\xi(N)$, where $N$ is the number of fibers in a bundle, so far as is known, has not been studied analytically or numerically.

In this paper, we will study analytically and numerically the dependences $\xi(\Delta f)$ and $\xi(N)$ for a uniform distribution of breaking thresholds in a fiber bundle, and investigate the reason for evolution of $\xi(\Delta f)$ values in the range of $-3 \div -5/2$.

In section 1, the simplest of known FBM models [4-6] will be described. In section 2, the distribution of burst probability of fiber bursts on each loading step will be obtained analytically. In section 3, we will compare the results of numerical simulation of FBM model and calculation according to analytical dependences obtained in section 2, a general view of dependence $\xi(\Delta f)$ and $\xi(N)$ will be given. The last section will be dedicated to numerical investigation of the type of distributions of fiber bursts on each loading step. It will be shown that different $\Delta f$ generate different types of above step distributions that leads to the evolution of $\xi(\Delta f)$ behavior.

**1. Fiber bundle model**

Numerous FBM models can be divided into different classes and types [1-6] according to the manner of load apply, the type of distribution of breakdown thresholds, etc. The simplest of known model is schematically shown in Fig.1.

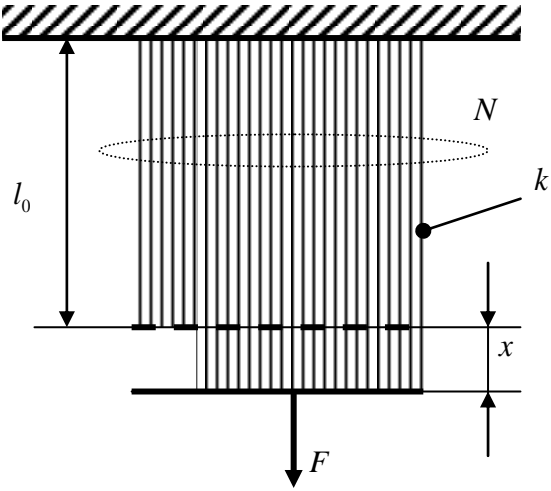

Fig. 1 Schematic of a fiber bundle in FBM model.

A bundle of $N$ identical fibers is stretched, each fiber having length $l_0$ in the free state. On stretching, a fiber obeys the Hooke law $F = -kx$ and has stiffness $k = 1$. The bundle is stretched by force $F$, with the bundle elongation $x$ (see Fig.1).

The distribution of fiber breakdown thresholds is distributed uniformly: $p(x) = 1, x \in [0..1]$, i.e. the whole bundle will be definitely broken if it is stretched by $x = 1$. The number of broken fibers at bundle stretched by $x$ will be $P(x) = x, x \in [0..1]$, unbroken fibers - $1 - P(x) = 1 - x, x \in [0..1]$.

Hence, the force stretching of fiber bundle by $x$ in the FBM model under study can be expressed as follows

$$F = N(1-x)x. \qquad (1)$$

Increase of a force applied to the bundle can cause avalanche burst of fibers. With the next burst of a single fiber it may turn out that the bundle stiffness will be reduced so much that it will be stretched enough for the burst of the next fiber, etc. The force necessary for the burst of the whole bundle can be found from (1) and equals to

$$F_{max} = N/4. \qquad (2)$$

It was shown analytically [10, 11], and then obtained by numerical simulation (see, for instance, references in [3-4]) that dependence $D(\Delta)$ - the number of above described avalanches on their size $\Delta$ - should follows a power law distribution

$$D(\Delta) \sim \Delta^{\xi}. \qquad (3)$$

Parameter $\xi$ - the integral characteristic of investigated FBM model – describes the behavior of fiber bursts at sufficiently large $\Delta$.

Experimental studies on the breakdown of various physical objects, such as snow [13], biological callogents [14], when predicting earthquakes [2,6] and others [5] demonstrate behaviour similar to that shown by the above described FBM model. There was also observed a deep relation between FBM model and the models of self-organized criticality [15] and biased random walk processes [16].

**2. Continuous and discrete loading**

In the analytical calculation of parameter $\xi$ in case of "low" [11] and "large" [12] $\Delta f$ the idea of stepwise loading was used. Thus, in the first case $\Delta f \to 0$, and it became possible to write as integrals the sums resulting from calculations and to obtain their asymptotic.

Let us consider stepwise loading of the bundle with a force step $\Delta f$. During the $m$-th step the force $f(m)$ is

$$f(m) = m \cdot \Delta f . \tag{4}$$

The total number of steps $M$, $m = 1..M$ will be determined by loading step $\Delta f$ and maximum force $F_{\max}$ (2) applied to the bundle

$$M = \frac{F_{\max}}{\Delta f} = \frac{N}{4 \cdot \Delta f} . \tag{5}$$

During each loading step the force $f(m)$ satisfies (3), and bundle elongation $x(m)$ corresponding to this force can be found with regard to (1) and (5) in the form:

$$x(m) = \frac{1}{2} - \sqrt{1 - \frac{4f(m)}{N}} = \frac{1}{2} - \sqrt{1 - \frac{m}{M}} , \tag{6}$$

During each loading step the bundle will be elongated by the value $\Delta x(m)$

$$\Delta x(m) = x(m+1) - x(m) = x(m) + x'_f(m)\Delta f - x(m) = \frac{\Delta f}{N\sqrt{1 - 4f(m)/N}} \tag{7}$$

Correspondingly, during the $m$-th step, with regard to uniform breakdown threshold distribution, $N\Delta x(m)$ fibers will be broken

$$N\Delta x(m) = \frac{\Delta f}{\sqrt{1 - m/M}} . \tag{8}$$

Taking into account that the number of breaks during the $m$-th step should be distributed according to the Poisson distribution, it is possible to make an expression for the burst probability of $k$ fibers during the $m$-th step $q(k,m)$

$$q(k,m) = \frac{(N\Delta x(m))^k}{k!} e^{-N\Delta x(m)}. \tag{9}$$

Hereinafter, step distribution of bursts is called the distribution of fiber bursts on each separate loading step, for instance, $m$-th bundle loading step.

According to definition, $\sum_{k=0}^{\infty} q(k,m) = 1$ should be met for any $m$, thus, for the entire process of bundle loading we will obtain, summing over all loading steps, $m = 1..M$

$$\sum_{m=1}^{M} \sum_{k=0}^{\infty} q(k,m) = M. \tag{10}$$

Let us define $Q(k)$ as follows

$$Q(k) = \frac{1}{M} \sum_{m=1}^{M} q(k,m). \tag{11}$$

Changing the order of summation, (10) can be rewritten as follows

$$\sum_{k=0}^{\infty} Q(k) = 1. \tag{12}$$

Thus, the value $Q(k)$ is the probability of breaking $k$ fibers in the process of stepwise bundle loading during for all bundle loading steps $m = 1..M$. $Q(k)$ depends on $\Delta f$ applied to the bundle in the same way for both continuous and discrete loading. $Q(k)$ corresponds to $D(\Delta)$ in the description of FBM (in the Introduction).

$$Q(k) \sim D(\Delta)|_{k=\Delta}. \tag{13}$$

The consequence of (12) is the fact that $D(\Delta)$ should not be normalized for the number of fibers in a bundle $N$, as is customary, but for the number of bundle loading steps $M$ / In this case the value $D(\Delta)/M$ can be interpreted not as a certain frequency of events, but as a probability (in the general sense of this concept) of appearance of burst avalanche of size $\Delta$ on loading of fiber bundle of size $N$. When performing numerical modelling of FBM presented in following sections we also calculated the value $\mathbf{D} = \sum_{k=0}^{\infty} (D(\Delta)/M)$, the analogue of formula (12), in all the calculations without exception $\mathbf{D} = 1.000 \pm 0.001$.

Certainly, behavior of $D(\Delta)$ and calculated value of model parameter $\xi$ are not affected by such normalization.

## 3. Burst distribution simulation

Dependences $Q(k)$ were calculated acc. to (11) numerically for different $\Delta f$. Selection of force step value $\Delta f$ for the subsequent calculation and numerical simulation was determined by the fact that parameter $\xi$ should cover the entire spectrum of its values $\xi = -3 \div -5/2$. The values were selected as follows: $\Delta f = 0.1, 1.0, 10.0, 30.0$ (see later Fig.4 and the comments to it).

Fig. 2 shows the plots of calculated dependences (11) for $\Delta f = 0.1, 1.0, 10.0$

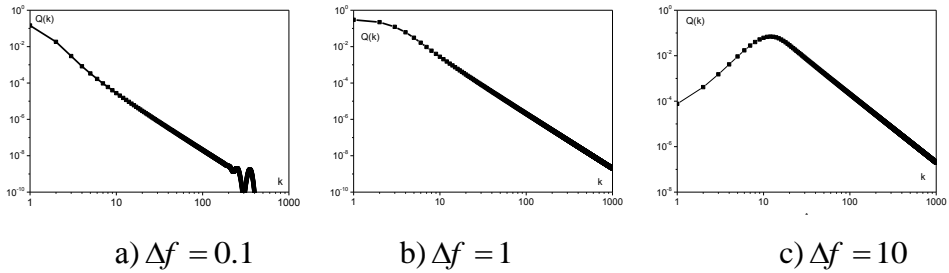

a) $\Delta f = 0.1$     b) $\Delta f = 1$     c) $\Delta f = 10$

Fig.2 The plots of dependence $Q(k)$ for different values of $\Delta f$.

Obtained in the numerical simulation of FBM, these dependences are given in Fig.3 for the same values of $\Delta f = 0.1, 1.0, 10.0$, the data was obtained by averaging over 100 realizations of a bundle with $N = 10^{10}$ fibres.

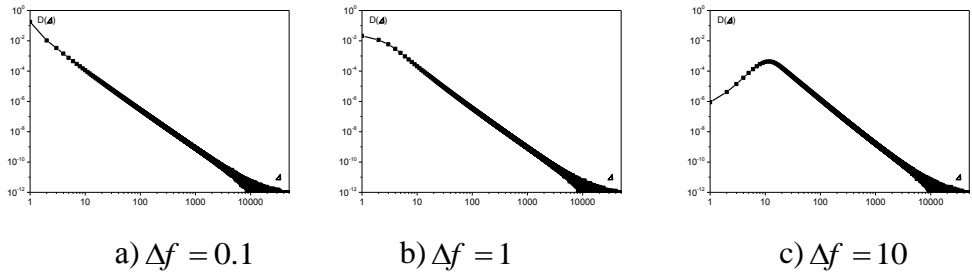

a) $\Delta f = 0.1$     b) $\Delta f = 1$     c) $\Delta f = 10$

Fig. 3 The plots of dependence $D(\Delta)$ for different values of $\Delta f$.

From comparison of Fig. 2 and 3 it can be seen that distribution (11) provides for a very detailed description of the numerical simulation. The initial portion of dependence at low values of $\Delta < \Delta_{min}$: on this portion at a relatively large step $\Delta f$ several fibres are broken, and single (double, ternary) bursts are few (Fig. 2c, 3c), or at low $\Delta f$ single bursts are predominant in the distribution (Fig. 2a, 3a). The final portion - large $\Delta$, $\Delta > \Delta_{max}$: numerical simulation (Fig.3) gives insufficient data for the adequate statistics, and in the calculations according to (11) one should use special methods to preserve the accuracy of the results obtained. These portions are impossible to be used for calculation of $\xi$.

Thus, only the medium part of the dependence (Fig.3), having no artifacts, enough data for statistical analysis, and where the slope of dependence $D(\Delta)$ remains constant, can be used for the calculation of $\xi$ from the results of numerical simulation.

Calculation of $\xi$ obtained from the results of calculation of $Q(k)$ according to (11) (see Fig.2) for all values of $\Delta f$ yields the same result: $\xi = -3.0$, with the numerical simulation $\xi$ takes on the values in the range $\xi = -3 \div -5/2$

The boundaries between these portions are intuitively obvious, but each researcher determines them himself, and the information on the way they are determined is generally unavailable. Changing these boundaries affects the calculated value of $\xi$, particularly in the region close to $\Delta_{min}$ where the absolute values of $D(\Delta)$ are $6 \div 8$ orders higher than in the region $\Delta_{max}$, and one can see some change of the distribution slope $D(\Delta)$ (see Fig 3a, 3b) in the region of $\Delta = 10 \div 50$. Since behaviour at large $\Delta$ is of primary interest, the respective region of the middle portion of dependence $D(\Delta)$ was used for calculations.

The data further given in Fig.4 were obtained with the following boundaries:

$$\Delta_{max} = \min_{\Delta}(D(\Delta) = 1)/2$$
$$\Delta_{min} = \Delta_{max}/10$$
(14)

Then the results of numerical simulation of FBM model are given, made for bundles with different number of fibers $N = 10^5 \div 10^{10}$.

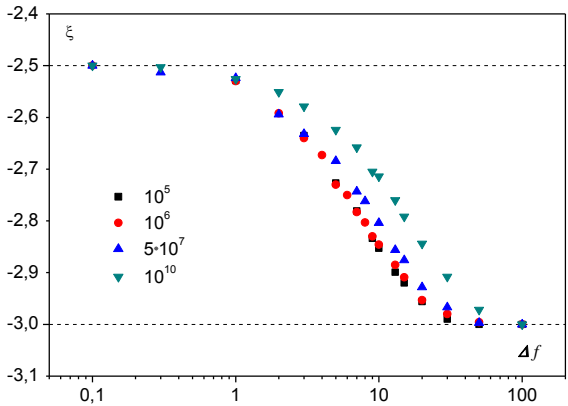

Fig.4 Dependence $\xi(\Delta f, N)$ for different $N$ calculated from the data of numerical simulation of FBM model.

Fig. 4 graphically represents parameters $\xi$ of FBM model calculated by least squares method from the data of numerical simulation of FBM model from the data restricted by conditions (14). At "low" $\Delta f < 0.1$ $\xi = -5/2$ [10,11] and "large" $\Delta f > 30$ $\xi = -3.0$ [12]. Note that as long as $q(k,m)$ (8) does not depend on $N$, it should be expected that $Q(k)$ will not depend on $N$ either. Indeed, from Fig. 4 it is seen that at "low" and "large" $\Delta f$ there is no dependence of $\xi$ on $N$, and in the transition region this dependence can be described as weak.

The results presented in Fig. 4, determined the set $\Delta f = 0.1, 1.0, 10.0, 30.0$ for calculations of FBM model. The values of $\Delta f = 0.1, 1.0$ are on the boundary $\xi = -5/2$, $\Delta f = 30.0$ in the region $\xi = -3$, and $\Delta f = 10.0$ in the transition region.

### 4. Step distributions of fiber bundle bursts

Calculation of parameter $\xi$ is certainly possible for the distributions of $Q(k)$ (Fig.2), obtained from calculation according to (11). Despite the fact that $\Delta f$ has assumed all the values from the above, the calculated value of slope for dependence $Q(k)$ has always been equal to $\xi = -3.0$. Thus, the employed ratio for $q(k,m)$ based on the Poisson distribution should with necessity give the same result $\xi = -3.0$, irrespective of the $\Delta f$ value. To put it differently, the type of distribution $q(k,m)$ determines the type of $Q(k)$ and, as a consequence, the value of parameter $\xi$.

Numerical simulation allows to find experimentally the step distributions $q_s(\Delta,m)$, making no preliminary assumptions on its form. For this purpose it is necessary, having assigned oneself with step number $m$, or, which is more convenient, the ratio $m/M$, to make a sufficient set of bundle bursts realizations, to reconstruct the form $q_s(\Delta,m)$. In practice, the values $m/M = 0.0, 0.1, 0.2 ... 1.0$ were selected, and moreover, to increase the data involved, we use the data for bursts $1 \cdot 10^{-3} M$ steps nearly each $m/M$-th loading step. Dependence $q_s(\Delta,m)$ was normalized for the maximum value $q_s(\Delta, m/M = 0)$.

Fig.5 represents the respective dependences $q_s(\Delta,m)$ for a bundle of size $N = 5 \cdot 10^7$ fibres obtained according to the results of 5000 realizations.

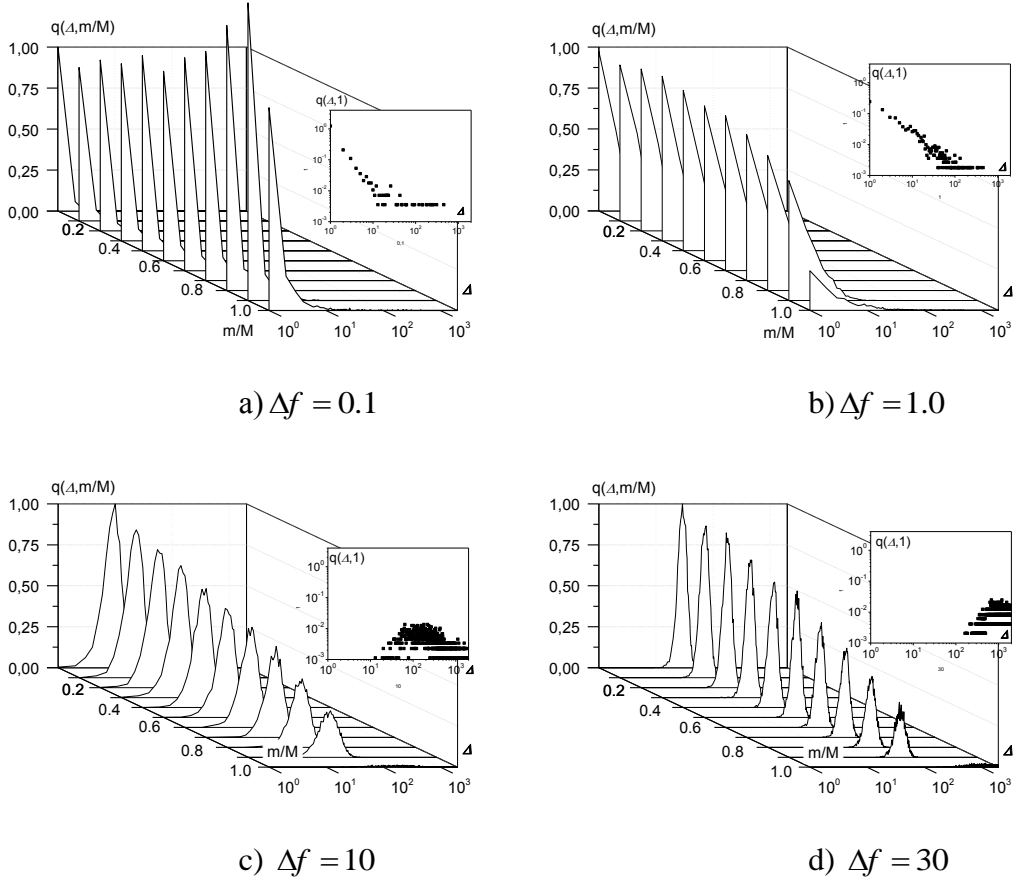

a) $\Delta f = 0.1$      b) $\Delta f = 1.0$

c) $\Delta f = 10$      d) $\Delta f = 30$

Fig. 5 A set of step distributions $q_s(\Delta, m)$ for different $\Delta f$ and $m/M$ (the values are indicated in the figure, the scale of all the figures is identical). On the insets – dependence of $q_s(\Delta, m/M = 1)$ on $\Delta$ in a log-log scale.

The data set presented in Fig. 5 provides the answer to the question on the reasons for evolution of FBM model parameter value $\xi(\Delta f)$.

Consider first $\Delta f = 30$ (Fig. 5d). During all bundle loading steps the step distribution corresponds to the Poisson distribution (or at least is very close to it). Thus, for instance, maxima on the obtained step distributions $q_s(\Delta, m)$ correspond to value (8) $N\Delta x(m)$ - the number of fibers broken on step $m$. The analytical (11) and numerical simulations coincide: $\xi(\Delta f > 30) = -3.0$.

Step distributions for $\Delta f = 10$ (Fig. 5c) and even $\Delta f = 1.0$ (Fig. 5b) also look similar to the Poisson distribution $D(\Delta)$. It should be emphasized that the initial portion of dependence $D(\Delta)$ in the region $\Delta = 5 \div 50$ (see Fig. 3a, 3b) is somewhat more sloping than the remaining part of distribution $D(\Delta)$, and it is exactly in this region that the "Poisson" maxima of stepwise distributions are located (see Fig. 5b,c), in this region $\xi \approx -3.0$. For large values of $\Delta > 50$ the distribution $D(\Delta)$ is accumulated on the tails of step distributions $q_s(\Delta, m)$, where only single bursts occur, they are scattered over the entire range of possible dimensions such that in the scale of Fig. 5b and 5c they are not seen. The reason for a change in $D(\Delta)$ type and evolution of $\xi(\Delta f)$ values lies in a modification of step distributions $q_s(\Delta, m)$ at large $\Delta$.

Changes in the numerically obtained step distributions $q_s(\Delta, m)$ are most distinctly seen at $\Delta f = 0.1$. At all loading steps $N\Delta x(m) \leq 1$, the entire distribution $D(\Delta)$ is accumulated on the tails of step distributions, the inset of Fig. 5a shows the distribution $q_s(\Delta, 1)$ on a log-log scale, so that single avalanches of large bursts can be seen. Therefore, at $\Delta f < 0.1$ a step distribution is completely transformed into another, non-Poisson tail distribution, and $\xi(\Delta f)$ ceases evolution, now it again becomes constant: $\xi(\Delta f \to 0) = -5/2$.

An example of step distribution approximating $q_s(\Delta, m)$ observed in numerical simulations (Fig.5a,b) can be, for instance, the following type of distribution.

$$q_{ex}(k, m) = \left(\frac{m}{M}\right)^{\gamma+1} \frac{1}{(k - m/M)^{\gamma}} \qquad (15)$$

The calculated distribution $Q_{ex}(k)$ according to proposed rule for $q_{ex}(k, m)$ at $\gamma = -2.5$ yields the value of parameter $\xi = -5/2$. Certainly, there are no special prerequisites, except for "appearance", for asserting that at $\Delta f < 0.1$ exactly this distribution holds, but as approximation to true distribution the description of which and the respective relation is not available so far, formula (15) is phenomenologically correct. The use of (15) brings into accord the results of analytical calculations and the numerical simulations.

It should be also noted that with the use of $\gamma = -3.0$ as a result for $Q_{ex}(k)$ we obtain $\xi = -3.0$, which on the one hand suggests the possibility of existing other types of step distribution resulting in $\xi = -3.0$, but in our case it is a wrong type of stepwise distribution, such approximation will not be phenomenologically correct approximation of the observed step distributions.

**Conclusion**

In this paper, within the probability-theoretic approach, a continuous and discrete loading of fiber bundle with a equal load sharing was studied. The bundle burst distribution was obtained as a sum of step burst distributions, i.e. distributions of fiber bursts on each loading step. Such an approach enabled a detailed analytical investigation of this distribution, as well as to apply the resulting relationships for a numerical simulations of distribution and its parameters. The analytical distributions and their parameters corresponded qualitatively and partly quantitatively to the results of numerical simulation of FBM model, and observed quantitative discrepancies allowed to determine the phenomenological reason for a change in bundle burst distribution in going from continuous to discrete loading and the evolution of this distribution parameter in the range from $-5/2$ to $-3$.

Numerical simulation of bundles with the number of fibers to $10^{10}$ was done. As it turned out, the number of fibers in a bundle has no essential effect on the type of bundle burst distribution, which corresponds to theoretical predictions. Thus, increasing the number of fibers in a bundle can only increase the area convenient for calculation of distribution parameters, but will not change the type of distributions or their parameters.

In the analytical investigation of bundle burst distribution there was employed a very logical assumption of the Poisson distribution of step bursts. It was confirmed for a discrete loading of the bundle, but with a continuous loading the analytical and numerical calculations did not fit each other.

The peculiarity of numerical simulations, i.e. the possibility to get insight into the details of running process, made it possible to determine that under a continuous loading the kind of step distributions is different from the Poisson one, at least for large-sized bursts. In this paper we have not determined the analytical type of the new distribution and the reasons leading to a change. The authors believe that "logical" determination of the step distribution function or the mechanism and method for transformation of step distributions would help to elucidate the processes occurring in the composite not only in simulation, but also in carrying out a full-scale physical experiment.

The choice of adequate load step in numerical simulation for the purpose of its further comparison to the results of full-scale physical experiment seems to be an interesting and nontrivial problem to the authors. Still more interesting seems to be a problem of fiber bundle burst on loading it in such a way that the choice of loading step is excluded.